\documentclass[12pt]{article}

\usepackage{epsfig}

\begin{document}

\title{\ The Stationary Points and Structure of High-Energy Scattering Amplitude}

\author{A.P.~Samokhin and V.A.~Petrov \\
\textit{A.A. Logunov Institute for High Energy Physics}\\
\textit{of NRC ``Kurchatov Institute''}\\
\textit{Protvino, 142281, Russian Federation}}

\date{}

\maketitle

\begin{abstract}
The ISR and the 7 TeV LHC data indicate that the differential cross-section of elastic
proton-proton scattering remains almost energy-independent at the
transferred momentum $t\approx - 0.21\, \mathrm{GeV}^{2}$ at the level
of $\approx 7.5 \mathrm{\ mb}/\mathrm{GeV}^{2}$. This property of $ d\sigma/dt$ (the ``first''
stationary point) appears due to the correlated growth of the total cross-section and
the local slope parameter and can be expressed as a relation between the latter quantities.
We anticipate that this property will be true up to 13 TeV.
This enables us to normalize the preliminary TOTEM data for $ d\sigma/dt$ at 13 TeV and
$ 0.05 < |t| < 3.4\, \mathrm{GeV}^{2}$ and predict the values of $ d\sigma/dt$ at this energy. 
These data give an evidence of the second stationary point
at $t\approx - 2.3\,\mathrm{GeV}^{2}$ at the level of $\approx 33 \mathrm{\ nb}/\mathrm{GeV}^{2}$.
The energy evolution of $ d\sigma/dt$ looks as if the high energy elastic scattering amplitude is a sum of two similar terms.  
We argue that the existence of the two stationary points and the two-component structure
of the high energy elastic scattering amplitude are general properties for all elastic processes.
\end{abstract}


\section{Introduction}

The simplest, at first sight, hadron-hadron elastic scattering process at low transferred momentum
is in reality one of the most complicated problems of high energy physics. The properties of the 
elastic scattering amplitude in this diffraction region are dominated by unknown, essentially
nonperturbative properties of the fundamental strong interactions. This is why we do not have so far
an adequate model for the soft hadronic phenomena and why at every generation of accelerators 
one reveals some new \textit{unexpected} properties of the hadron-hadron elastic scattering.

In the present paper we discuss one of such new unexpected properties of the elastic differential 
cross-section. In Section 2 it is shown that the ISR and the 7 TeV LHC data give an evidence of 
the stationary point of $ d\sigma/dt$ in the forward peak region. We analyse the nature of this 
new scaling property which leads to a specific relation between the total cross-section and the
mean value of the local slope. In Section 3 we estimate the value of energy at which this new
scaling will be broken. Suggesting that the stationarity persists up to 13 TeV we normalize
the preliminary 13 TeV TOTEM data and predict the values of $ d\sigma/dt$ at this energy. Comparing in
Section 4 the behaviour of $ d\sigma/dt$ in the region beyond the second maximum at the ISR and LHC
energies we have got an evidence of the second stationary point. The existence of two shrinking with
energy diffraction cones motivates the two-component structure of the high energy pp elastic 
scattering amplitude. We argue that the latter are general properties of elastic scattering.
A brief summary and discussion are given in Section 5.

\section{The first stationary point}

Comparing the differential cross-section for pp elastic scattering
$ d\sigma(s,t)/dt$ at the ISR energies [1] with the 7 TeV LHC data [2]
in the forward peak region one can observe that the shrinkage of the
diffraction cone in this energy range goes in some specific way.
At any fixed value of the transferred momentum $(-t) \in [0,\, 0.21)$ GeV$^{2}$
the differential cross-section grows with energy, while for $(-t) \in (0.21,\, 0.53)$ GeV$^{2}$
it decreases. At $t\approx - 0.21$ GeV$^{2}$
the differential cross-section remains almost energy-independent
at energies from the ISR up to 7 TeV (see Fig. 1). Thus, there is an evidence of
\textit{a stationary point} [3] of the differential cross-sections $(t_{\ast}, \sigma_{\ast})$ :
\begin{equation}
 t_{\ast} \approx - 0.21 \pm 0.01 \,\, \mathrm{GeV}^{2} ,\,\,\,  \sigma_{\ast}(s)\equiv \frac{d\sigma(s,t_{\ast})}{dt} \approx 7.5 \pm 0.5 \,\,
 \mathrm{\ mb}/\mathrm{GeV}^{2},  
\end{equation}
where $ t_{\ast}$ is fixed (energy independent) and $ \sigma_{\ast}(s)$ practically does not vary at energies from the ISR up to 7 TeV.
\begin{figure}[t]
\centering
\includegraphics[height=8.5cm]{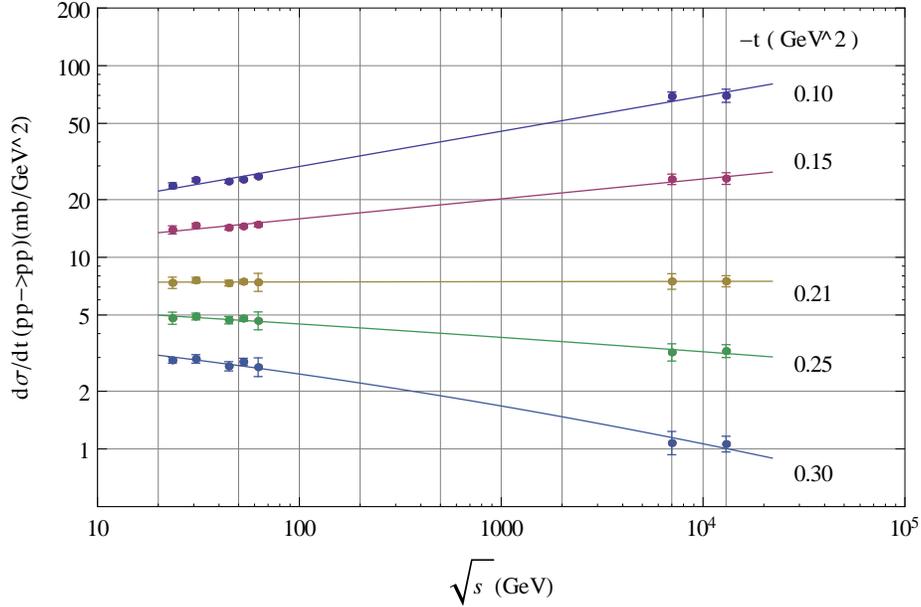}
\caption{Energy evolution of the differential cross-section for pp
elastic scattering at fixed values of transferred momenta in the vicinity of $ t_{\ast}$. The ISR and 7 TeV TOTEM experimental
data are from Refs. [1], [2]. The points at $ \sqrt{s} = 13$ TeV are the TOTEM preliminary data [11] in our normalization (11),
see Table 2.}
\end{figure}

Unfortunately, we do not have experimental data for pp elastic scattering at energies between
the ISR and 7 TeV. But the 2.76 TeV TOTEM data for $ d\sigma/dt$ will be available soon and can
be compared with the predictions of Fig. 1. Let us note that we cannot use the SPS and Tevatron
p$\bar{\mathrm{p}}$ data in this analysis because the value of the odderon impact is unknown \textit{a priori}.
The p$\bar{\mathrm{p}}$ elastic scattering data must be analysed separately.

The mathematical origin of the discussed phenomenon is similar to that of geometrical scaling [4]
in the ISR energy range where the growth of the elastic and total cross-sections and  
the growth of the forward slope, $B(s)$, approximately compensate each other in the ratios
\begin{equation}
   \frac{\sigma_{el}(s)}{\sigma_{tot}(s)}\approx const,\,\,\,\,\,\,\,\,  \frac{\sigma_{tot}(s)}{B(s)}\approx const. 
\end{equation}
At low energies (up to the ISR) these ratios decrease while at higher energies (starting from the SPS)
they grow with the energy growth. In this way the relations (2) are a ``transitory'' property of these ratios.

The behaviour of the differential cross-section 
\begin{equation}
   \frac{d\sigma(s,t)}{dt} = \sigma_{0}(s)exp[\int\limits_{0}^{t}dt^{'}B(s,t^{'})]  
\end{equation}
at fixed $t$ in the forward peak range is defined by the energy evolution of 
\begin{equation}
\sigma_{0}(s)\equiv \frac{d\sigma(s,t)}{dt}|_{t=0} = \frac{\sigma_{tot}^{2}(s)(1+\rho^{2}(s))}{16\pi}\,,\,\,\,\,\,\, 
\rho(s) = \frac{ReT(s,0)}{ImT(s,0)} 
\end{equation}
and of the local slope
\begin{equation}
   B(s,t)= \frac{d}{dt} ( \ln[\frac{d\sigma(s,t)}{dt}]) .
\end{equation}
In the forward peak region the local slope is an increasing function of energy at fixed $t$, therefore the exponential
factor in Eq. (3) decreases with energy at any fixed $t$. 
At low energies (up to the ISR) the factor $ \sigma_{0}(s)$ decreases with energy, therefore
$ d\sigma(s,t)/dt$ decreases with energy at any fixed $t$ also.

Starting from the ISR where $ \sigma_{tot}(s) $ begins to grow the growth of $ \sigma_{0}(s)$
competes with the growth of the slope parameter. At low fixed values of $ |t| < |t_{\ast}|$
the growth of $ \sigma_{0}(s)$ dominates and, as a result, $ d\sigma(s,t)/dt$ grows (see Fig. 1).
At fixed $ |t| > |t_{\ast}|$ the growth of $ B(s,t)$ dominates and, as a result, 
$ d\sigma(s,t)/dt$ decreases with energy (see Fig. 1). At $ t = t_{\ast}$ the growth of $ \sigma_{0}(s)$
is compensated by the growth of the slope in Eq. (3) and we have the stationary point (see Eq. (1) and Fig. 1) in
the whole energy range from the ISR up to the LHC.

As well as the geometrical scaling (and the Bjorken scaling of the structure function $ F_{2}^{pp}(x,Q^{2})$ at the ``pivot'' point [5]) 
this new scaling has a transitory character. Indeed, if $ \sigma_{tot}(s)\rightarrow\infty$ at $ s\rightarrow\infty$
and if according to Fig. 1 and Eq. (1) $d\sigma(s,t)/dt = \pi |A(s,t)|^{2}$ goes to infinity at any fixed $t$ from the
\textit{fixed range} $ t_{\ast} < t \leq 0$ then the partial wave $ a_{0}(s)$ of the scattering amplitude $ A(s,t)$
\begin{equation}
 a_{0}(s) = \frac{1}{4} \int\limits_{-s}^{0}dt^{'}A(s,t^{'}) \approx \frac{1}{4} \int\limits_{t_{\ast}}^{0}dt^{'}A(s,t^{'}) = 
 \frac{1}{4} A(s,\tilde{t})|t_{\ast}|,\,\,\,\tilde{t} \in (t_{\ast},0) 
\end{equation}
goes to infinity at $ s\rightarrow\infty$ also since $ |t_{\ast}|$ is fixed.
But the latter is impossible because due to the unitarity the partial waves must be bounded.
Thus, $ \sigma_{\ast}(s) = d\sigma(s,t_{\ast})/dt $ from Eq. (1) must begin to decrease at some high energy. In general,
it is naturally to suppose [6] that $ d\sigma(s,t)/dt$ decreases at $ s\rightarrow\infty$ for any \textit{fixed} (energy independent) $ t < 0$.
So, the growth of $ d\sigma/dt$ with energy at any fixed $ 0 < |t| < |t_{\ast}|$ (see Fig. 1) must be changed by a decrease starting from some 
high energy. The validity of Eq. (1) and Fig. 1 up to 7 TeV means, by the way, that the LHC energies are far away from the asymptotics. 

Taking use of the mean value theorem for the integral in Eq. (3) we have 
\begin{equation}
 \frac{d\sigma(s,t)}{dt} = \sigma_{0}(s)exp(\tilde{B}t),   
\end{equation}
where $ \sigma_{0}(s)$ is given by Eq. (4) and $\tilde{B}$ is the local slope $ B(s,t)$ in some inner point $ \tilde{t}$
\begin{equation}
 \tilde{B}\equiv B(s,\tilde{t}),\,\,\, \tilde{t} \in [t,0],\,\,\, \tilde{t} = \tilde{t}(t,s). 
\end{equation}
In particular, at $ t =t_{\ast} $ we have
\begin{equation}
\sigma_{\ast}(s) = \sigma_{0}(s)exp(\tilde{B_{\ast}}t_{\ast}),\,\,\,\,\, \tilde{B_{\ast}}\equiv B(s,\tilde{t}_{\ast}),\,\,\,\,\, 
\tilde{t}_{\ast} \in [t_{\ast},0].  
\end{equation}
Existence of the stationary point (Eq. (1)) means that at energies from the ISR up to 7 TeV 
\begin{equation}
\tilde{B_{\ast}}\equiv \frac{1}{(-t_{\ast})} \ln(\frac{\sigma_{0}(s)}{\sigma_{\ast}(s)}) \approx
9.52 \ln(\frac{\sigma_{tot} \sqrt{1+\rho^{2}}}{12.12(\mathrm{mb})})\, \mathrm{GeV}^{-2} ,
\end{equation}
that is the mean value of the local slope in the range $ [t_{\ast},0]$ is defined only by $ (\sigma_{tot} \sqrt{1+\rho^{2}})$ and increases 
approximately as $\ln(\sigma_{tot}(s))$. On the other hand, we can find the mean value of the local slope in the range $ [t_{\ast},0]$
from the experimental data Refs. [1], [2] for $ d\sigma/dt$ (according to the left-hand side of Eq. (10)) and compare this quantity $\tilde{B}_{exp}$ 
with $\tilde{B_{\ast}}$ from the rhs of Eq. (10). Table 1 shows that Eq. (10) is in accordance with the experimental data.
At the ISR energies $\tilde{B}_{exp}$ is less than $ B(s)$ because the local slope decreases with $|t|$ in the range $[0,|t_{\ast}|]$ [7], 
but at the LHC energies it may be equal or slightly more than $ B(s)$ because of the growth of the local slope [8] in the vicinity of
$t\approx - 0.2\, \mathrm{GeV}^{2}$.

So, we see that the stationarity of $ d\sigma/dt$ at $ t = t_{\ast} $ (Eq. (1) and Fig. 1) and, as a result, the relation (10)
are valid at energies from the ISR up to 7 TeV.
It would be very interesting to compare Eqs. (1), (10) with the data at intermediate (between the ISR and 7 TeV, in particular, at 2.76 TeV) 
energies and with the expected data at 13 TeV.
\begin{figure}[t]
\begin{center}
{\bf Table 1:}
{\small The mean value of the local slope $ \tilde{B_{\ast}}$ in the range $ [t_{\ast},0]$ computed from Eq. (10)
and the mean value of the local slope $\tilde{B}_{exp}$ computed from the experimental data Refs. [1], [2] for $ d\sigma/dt$.}
\\[2mm]
\includegraphics[height=2.0cm]{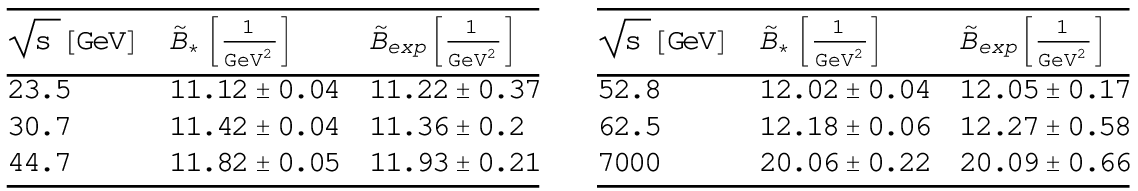}
\end{center}
\end{figure}

\section{Predictions for $ d\sigma/dt$ at 13 TeV}
As we have argued, the scaling property (1) must be broken at very high energies. Hence, the relation (10) as a direct
consequence of Eq. (1) will be true only up to some finite value of energy. To estimate this value of energy let us note
that the rhs of Eq. (10) must satisfy the unitarity lower bound [9] on the slope parameter: $ B(s) > \sigma_{tot}^{2}/18\pi \sigma_{el}$.
It is possible only at energies up to 150 TeV.
But a significant difference in the values of Eq. (10) and the often used approximate formula
$ B(s) \approx \sigma_{tot}^{2}/16\pi \sigma_{el}$ will be seen already at energies 20 $ \div$ 30 TeV (for these estimates we used
the fits to the $ \sigma_{tot}(s)$ and $ \sigma_{el}(s)$ from [10]). It means that the relation (10) (and therefore the scaling property (1))
will be approximately true at least up to 20 TeV. So, we can anticipate that 
\begin{equation}
\frac{d\sigma(s,t_{\ast})}{dt} = \sigma_{\ast}(s) \approx 7.5\pm 0.5\,\, \mathrm{\ mb}/\mathrm{GeV}^{2}\,\,  
\mathrm{at}\,\, \sqrt{s} = 13\, \mathrm{TeV}.
\end{equation}
If we assume that at 13 TeV $ \sigma_{tot}(s)\approx (109\pm 2) \mathrm{\ mb}$ [10] then according to Eq. (10) the 
mean value of the local slope at 13 TeV is $ \tilde{B_{\ast}} \approx (21.0\pm 0.5)\, \mathrm{GeV}^{-2}$.

The TOTEM Collaboration exhibits the preliminary unnormalized 13 TeV data [11] for the pp elastic differential cross-section  
in the $ 0.05 < |t| < 3.4\, \mathrm{GeV}^{2}$ region. The suggestion (11) enables us to normalize these data and have got the 
table of the values for $ d\sigma/dt$ at 13 TeV (see Table 2 and Fig. 1, Fig. 2). In particular, the value of $ d\sigma/dt$  
at the dip at 13 TeV is estimated as
\begin{equation}
 \frac{d\sigma}{dt}|_{dip} \approx 31.3\pm1.9\,\,  \mathrm{\mu b}/\mathrm{GeV}^{2},\,\,\,-t_{dip} \approx 0.483\pm0.011\,\, \mathrm{GeV}^{2}.  
\end{equation}
\begin{figure}[t]
\begin{center}
{\bf Table 2:}
{\small The digitizing of the preliminary unnormalized data [11] for the pp elastic differential cross-section 
at $ \sqrt{s} = 13$ TeV in our normalization (11).}
\\[2mm]
\includegraphics[height=11.0cm]{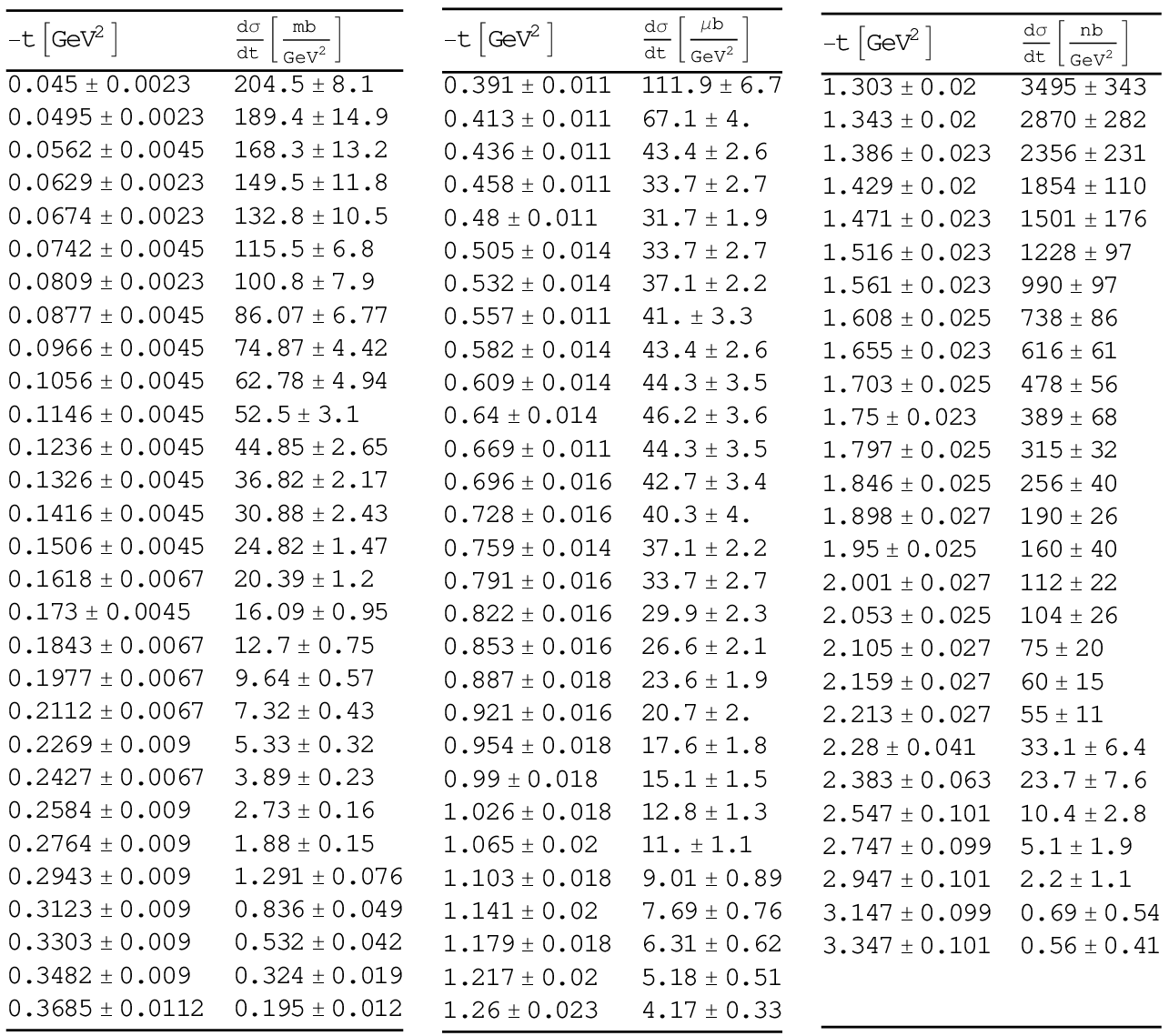}
\end{center}
\end{figure}
\begin{figure}[t]
\centering
\includegraphics[height=8.5cm]{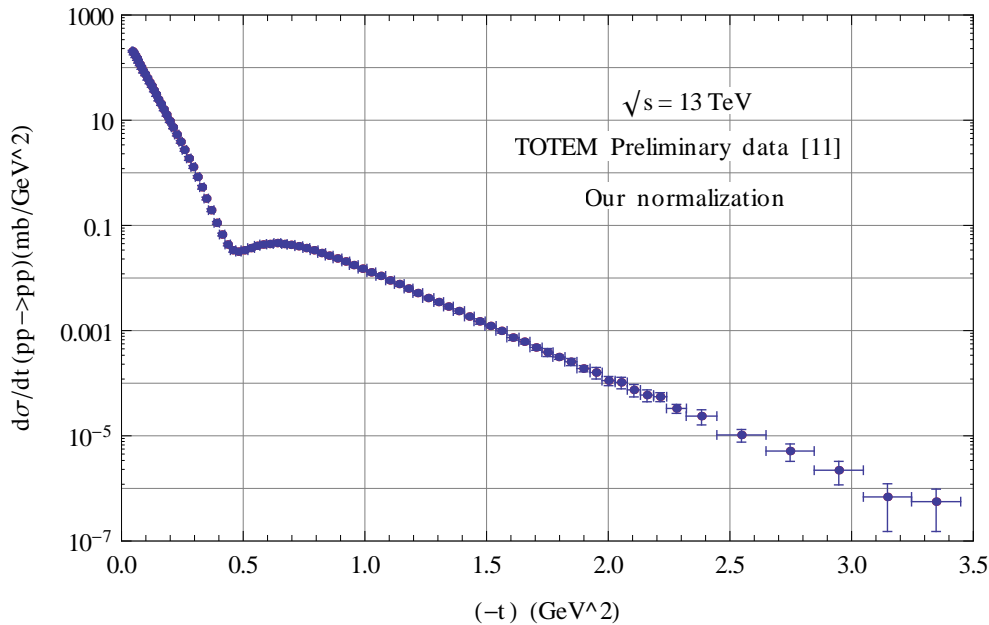}
\caption{The differential cross-section for pp
elastic scattering at $ \sqrt{s} = 13$ TeV [11] in our normalization (11), see the above Table 2.}     
\end{figure}

The TOTEM 13 TeV preliminary data [12] for $ d\sigma/dt$ in the region $ 0.002 < |t| < 0.02\, \mathrm{GeV}^{2}$ are in
line with our normalization (11).

\section{The second stationary point and structure of high energy scattering amplitude}
At Fig. 3 we have presented the differential cross-sections at the ISR and the LHC (7 and 13 TeV in our normalization) energies together.
Let us remind that the 7 TeV data [2] for $ d\sigma/dt$ are known only up to $ |t| \approx 2.4\, \mathrm{GeV}^{2}$, while 
the differential cross-section at 13 TeV for the first time in the LHC energy range is measured [11] up to $ |t| \approx 3.4\, \mathrm{GeV}^{2}$. 
It is a very important circumstance because due to these 13 TeV data (in our normalization) we see at Fig. 3
besides the stationary point at $t\approx - 0.21$ GeV$^{2}$ a clear-cut evidence of \textit{the second stationary point} at
\begin{equation}
t_{\ast \ast} \approx - 2.3\pm0.1\,\,\mathrm{GeV}^{2} ,\,\,\,  \sigma_{\ast \ast}(s)\equiv \frac{d\sigma(s,t_{\ast \ast})}{dt}
\approx 33\pm5\,\, \mathrm{\ nb}/\mathrm{GeV}^{2}.
\end{equation}
Indeed, at fixed values of transferred momenta $|t|$ beyond the second maximum of $ d\sigma/dt$ but less than $ |t_{\ast \ast}|$ 
the differential cross-section grows with energy, while for fixed $|t| \in (2.3,\, 3.4)$ GeV$^{2}$ it decreases with energy
(see Fig. 3, Fig. 4). At $ t = t_{\ast \ast} \approx - 2.3\,\mathrm{GeV}^{2}$ the differential cross-section remains almost
energy-independent at energies from the ISR up to 13 TeV (see Fig. 3, Fig. 4). In other words, we see that the energy behaviour
of $ d\sigma/dt$ in the vicinity of $ t = t_{\ast \ast}$ is exactly the same as the behaviour of $ d\sigma/dt$ in the diffraction cone
in the vicinity of $ t = t_{\ast}$. Moreover, Fig. 3 shows that there exists a certain correlation between shrinkage of \textit{the first and this 
second diffraction cones}: they have turned around their stationary points $t_{\ast}$, $t_{\ast \ast}$ approximately by 
the same angle when the energy grows from the ISR up to the LHC values. It is evident also, that as well as Eq. (1) the scaling property (13)
has a transitory character (due to the general arguments of Sec. 2).
\begin{figure}[t]
\centering
\includegraphics[height=8.5cm]{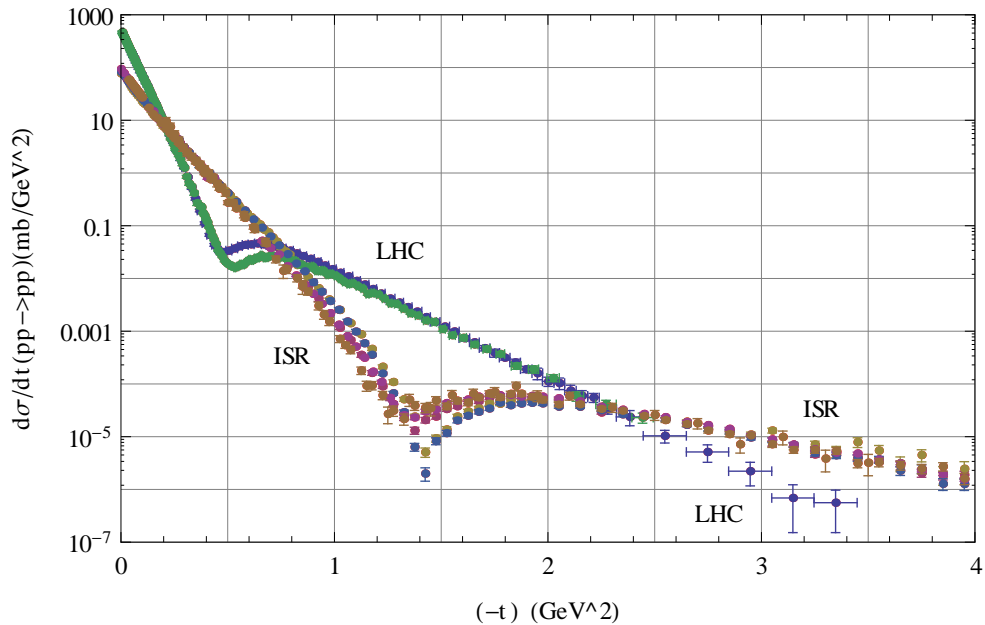}
\caption{
The ISR ($ \sqrt{s} = 23,30,44,53,62$ GeV$ $) and the LHC ($ \sqrt{s} = 7$ and $13$ TeV) differential cross-sections are shown together.
The experimental data are from Refs. [1], [2] and [11] in our normalization (11), see the above Table 2.}    
\end{figure}
\begin{figure}[t]
\centering
\includegraphics[height=8.5cm]{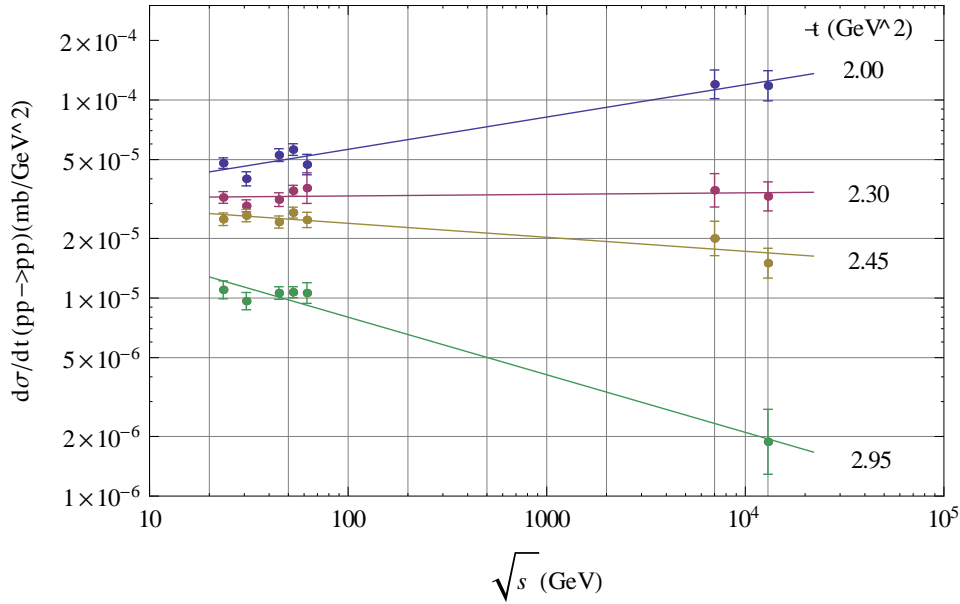}
\caption{Energy evolution of $ d\sigma(s,t)/dt$ at fixed values of transferred momenta in the vicinity of $ t_{\ast \ast}$. The experimental
data are from Refs. [1], [2] and [11] in our normalization (11), see the above Table 2.}  
\end{figure}

In the ISR energy range the differential cross-section $ d\sigma(s,t)/dt$ for fixed $|t| \in (2.3,\, 5.1)$ GeV$^{2}$ is approximately
energy-independent [13]. These data (as well as the ISR and Fermilab data in the $ 5 < |t| < 10\, \mathrm{GeV}^{2}$ region) are described
by the simple \textit{energy-independent} formula [14] $ d\sigma/dt \approx const |t|^{-8}$. Such energy-independent behaviour of $ d\sigma/dt$
for $ |t| \geq 3$ GeV$^{2}$ was expected up to the LHC energies, but the 13 TeV TOTEM data in the $ 2.3 < |t| < 3.4\, \mathrm{GeV}^{2}$ region
(see Fig. 3, Fig. 4) destroy these expectations. The differential cross-section instead reveals in this $t$-region the second shrinking 
with energy diffraction cone. It also means that a simple picture of perturbative exchange by bare gluons is not relevant at these values of $t$.

The existence of two similar diffraction cones of $d\sigma/dt = |T(s,t)|^{2}/16\pi s^{2}$ motivates the following natural assumption:
the high energy pp elastic scattering amplitude $T(s,t)$ in the region $ 0 \leq |t| \leq 3.4\,\mathrm{GeV}^{2}$ is a sum 
of two \textit{similar} terms
\begin{equation}
 T(s,t) = 4\pi s(A_{1}(s,t) + A_{2}(s,t)).
\end{equation}
The first term dominates in the forward peak region while the second one dominates in the region beyond the
second maximum of $ d\sigma/dt$. The energy evolution of the first diffraction cone (with the stationary point at $t = t_{\ast}$)
is determined by the behaviour of $ |A_{1}(s,t)|^{2} $ (the impact of $A_{2}(s,t)$ in the forward peak region is insignificant).
The mechanism of such behaviour of $ |A_{1}(s,t)|^{2} $ has been discussed in detail in Sec. 2. Accordingly, the energy evolution
of the second diffraction cone (with the stationary point at $t = t_{\ast \ast}$) is determined by the behaviour of $ |A_{2}(s,t)|^{2} $
(the impact of $A_{1}(s,t)$ in the region beyond the second maximum of $ d\sigma/dt$ is insignificant). It is naturally to assume
that the mechanism of energy behaviour of $ |A_{2}(s,t)|^{2} $ in the vicinity of $t = t_{\ast \ast}$ is similar to that of $ |A_{1}(s,t)|^{2} $
in the vicinity of $t = t_{\ast}$. It is the simplest (but not unique) possibility. 

If so, then the dip-bump structure in $ d\sigma/dt$ is due to the interference of these $ A_{1},\,A_{2} $ terms
\begin{equation}
 \frac{d\sigma(s,t)}{dt} = \pi [(|A_{1}| - |A_{2}|)^{2} + 2|A_{1}||A_{2}|(1 +\cos(\varphi_{1} - \varphi_{2}))],
\end{equation}
where $\varphi_{1}(s,t)$ and $\varphi_{2}(s,t)$ are the phases of $A_{1}(s,t)$ and $A_{2}(s,t)$ respectively. 
Just this interference structure of $d\sigma/dt$ in the ISR energy range initiated an appearance of the two-exponential
(with a relative phase) parametrization of the elastic scattering amplitude [15], [13]. Now, in the LHC era, 
the existence of two similar diffraction cones of $d\sigma/dt$ gives the additional arguments 
in favour of the conjecture that the high energy elastic scattering amplitude has a two-component structure (14).

As well known, the growth of $\sigma_{tot}(s)$ and hence (due to the MacDowell-Martin bound [9] $ B(s) > \sigma_{tot}^{2}/18\pi \sigma_{el}$) 
that of the slope $B(s)$ are  
universal properties of the hadron-hadron scattering. Therefore for any elastic process (pp, p$\bar{\mathrm{p}}$, $\pi$p, Kp ) there must exist
an energy region where the growth of $\sigma_{tot}(s)$ will be approximately compensated by the slope growth in $ d\sigma/dt$ at some
fixed value of the transferred momentum. So, the presence of a ``metastable'' stationary point in the forward peak region seems to be an
universal property also.
We anticipate that the second stationary point which lies beyond the second maximum of $ d\sigma/dt$ is a general feature
of all elastic processes also. Of course, the values of ($t_{\ast},\, \sigma_{\ast}$) and ($t_{\ast \ast},\, \sigma_{\ast \ast}$) may differ 
for different elastic processes, but the very fact of existence of the stationary points witnesses in favour of the two-component structure
of the high energy elastic scattering amplitude (14). It would be very interesting to check these anticipations for the p$\bar{\mathrm{p}}$,
$\pi$p and Kp elastic processes.

\section{Summary and discussion}
Let us sum up the above said. The ISR and the 7 TeV LHC data give an evidence of a stationary point Eq. (1). This scaling
property is equivalent to the connection (10) between the mean value of the local slope and $ \sigma_{tot}(s)$. We have discussed
the compensatory nature and the transitory character of Eqs. (1), (10). Supposing their validity at least up to 13 TeV we normalize the preliminary
13 TeV TOTEM data [11] and have got the table of the values for $ d\sigma/dt$ at 13 TeV. It enables us to get an evidence of a 
second stationary point in the region beyond the second maximum of $ d\sigma/dt$. We anticipate that the existence of the stationary 
points is a general property of all elastic processes and that it witnesses in favour of the two-component structure of the high energy elastic 
scattering amplitude (14).

In the ISR energy range it is observed an approximate geometrical scaling, when there are approximate relations (2) and $d\sigma/dt = R^{4}f(R^{2}t)$,
where $ R(s)$ is an effective interaction radius [4]. It was noticed [16], [17] \textit{en passant} that in this GS-regime the differential 
cross-section has a stationary point at $ t_{GS} = -2/B(s)$ (that is $ d(d\sigma/dt)/ds = 0$, when $ t = t_{GS}$). At the ISR energies it is not 
so bad, $ t_{GS} \approx - 0.2$ GeV$^{2}$, but at high energies this formula for a stationary point as well as the geometrical scaling itself are 
false. The differential cross-section instead remains approximately constant at $ t = t_{\ast} \approx -0.21$ GeV$^{2}$ and at energies from 20 GeV up 
to 20 TeV (see Fig. 1) due to the compensation of the growth of factor $ \sigma_{0}(s) \sim \sigma_{tot}^{2}(s)$ by the decrease of the
exponential factor in Eq. (3). Of course, this compensation has an approximate character and the value of $\sigma_{\ast}(s) = d\sigma(s,t_{\ast})/dt$ 
can slightly ``breathe'' when energy grows from the ISR up to the LHC, as it takes place, for example, for the COMPETE fit [18]. So, the ISR 
and LHC data reveal a new scaling property of $ d\sigma/dt$ and a new mechanism of shrinkage of the diffraction cone which are valid 
in the unprecedented wide range of energy. It enables to predict the behaviour of $ d\sigma/dt$ in the vicinity of $ t = t_{\ast}$ for 
the different values of energy (see Fig. 1). Moreover, it is enough to normalize a bulk of the preliminary data [11] and to predict the values 
for $ d\sigma/dt$ at 13 TeV in the entire $ 0.05 < |t| < 3.4\, \mathrm{GeV}^{2}$ region (see Fig. 3, Fig. 4).

The latter is very essential because so far we had for $ d\sigma/dt$ at large enough $ |t|$ only the ISR data, which are well described by
the energy-independent formula [14] $ d\sigma/dt \approx const |t|^{-8}$. The 7 TeV TOTEM data [19], [2] give the values of
$ d\sigma/dt$ only for $ |t| < 2.4\, \mathrm{GeV}^{2}$ (see Fig. 3). These data initiated a discussion of the possibility of new scattering 
mechanisms in the $|t| > 2.4\,\mathrm{GeV}^{2}$ region: remark on a possible change in the spectra at large $ |t|$ [20], the Orear-like 
$t-$behaviour of $ d\sigma/dt$ [21], the exponential $t-$behaviour in the modified Barger-Phillips model [22]. But only the 13 TeV TOTEM data [11] 
in the $ 2.3 < |t| < 3.4\, \mathrm{GeV}^{2}$ region proved definitely the energy dependence of $ d\sigma/dt$ for $|t| > 2.3\,\mathrm{GeV}^{2}$.
Moreover, the energy evolution of $ d\sigma/dt$ in the vicinity of $ t = t_{\ast \ast} \approx - 2.3\,\mathrm{GeV}^{2}$ is very similar to
the energy evolution of $ d\sigma/dt$ in the vicinity of $ t = t_{\ast} \approx - 0.21\,\mathrm{GeV}^{2}$ (see Fig. 3, Fig. 4 and Fig. 1). 
In other words, the ISR and LHC data in the region beyond the second maximum of $ d\sigma/dt$ reveal a second diffraction cone shrinking with energy
and demonstrating a stationary point at $ t = t_{\ast \ast}$. The diffractive character of $ d\sigma/dt$ in this region seems to rule out the dominance
of perturbative exchanges of a few non-interacting gluons which assumed in Ref. [14].
It is very interesting and unexpected property of the high energy behaviour of $ d\sigma/dt$.

The existence of two shrinking diffraction cones in the vicinity of $ t_{\ast}$ and $ t_{\ast \ast}$ gives the additional arguments in favour of
the two-component structure of the high energy elastic scattering amplitude, which was originally proposed to describe a dip-bump picture
in $ d\sigma/dt$ [15], [13]. Now there are many models of such type, for example: a modified two-exponential (with a relative phase) 
parametrization [22]; a model, where $ A_{1}$ is a dipole Pomeron and $ A_{2}$ is a dipole Odderon [23]; a model, where $ A_{1}$ is a soft Pomeron 
and $ A_{2}$ is a hard Pomeron [24]; a Pomeron pole plus grey disk model [25]. The similarity of the $ A_{1}(s,t)$ and $ A_{2}(s,t)$ properties
can help to construct an adequate model of the elastic scattering amplitude.

As we have seen above, the mechanism of shrinking of the first diffraction cone with a stationary point at $ t = t_{\ast}$ is a consequence
of the correlated growth of $\sigma_{tot}(s)$ and $ B(s)$. But the latter, as well as the dip-bump structure of $ d\sigma/dt$, is a general
property of the hadron-hadron scattering. Hence, we have good reason to anticipate that the stationary points in the first and second
diffraction cones will appear for all elastic processes and this, in turn, will witness in favour of the two-component structure of the
high energy elastic scattering amplitude (14). Evidently, the underlying reasons of that must be very general and fundamental.
In our opinion, such a form of amplitude may originate from a specific $s-t-u$ structure symmetry,
which was realized for the first time in the Mandelstam representation [26] and later in the dual models [27,28].

\textit{Note added in proof.} The TOTEM Collaboration has published [29] the luminosity-independent values of the proton-proton elastic, inelastic 
and total cross-section at $\sqrt{s} = 13$ TeV. This luminosity-independent value of $\sigma_{el} = 31.0 \mathrm{\ mb}$
was used to normalize the differential cross-section of elastic pp scattering at $\sqrt{s} = 13$ TeV in the $ 0 < |t| < 0.2\, \mathrm{GeV}^{2}$ 
region [30]. At Fig. 5 we see that the TOTEM data [30] lie slightly higher than the TOTEM preliminary data [11] in our normalization (11).
\begin{figure}[t]
\centering
\includegraphics[height=8.5cm]{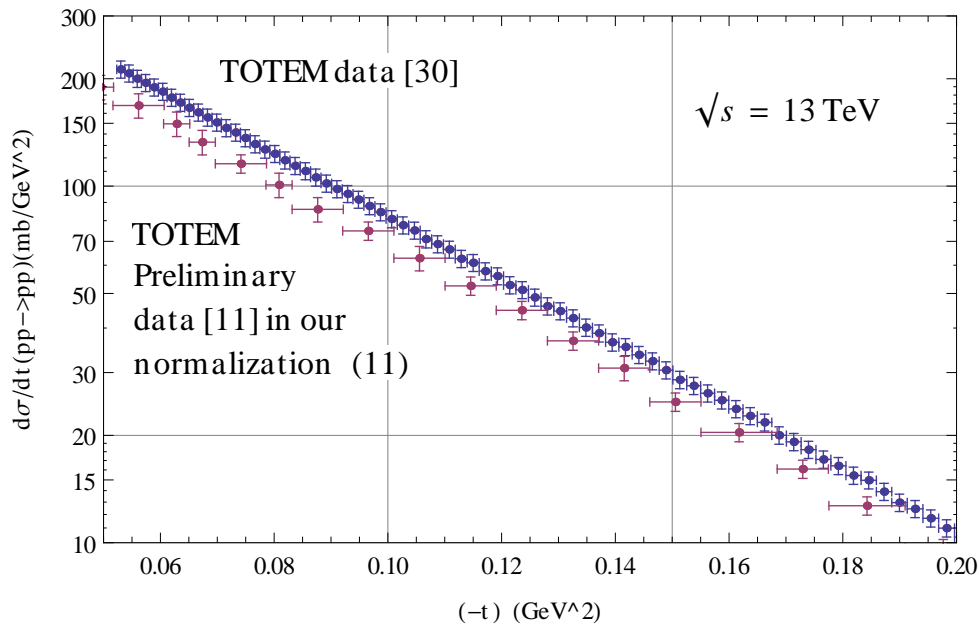}
\caption{
The differential cross-section of elastic pp scattering at $\sqrt{s} = 13$ TeV in the $ 0.05 < |t| < 0.2\, \mathrm{GeV}^{2}$ region.
The experimental data are from Refs. [30] and [11] in our normalization (11), see the above Table 2.}    
\end{figure}
Let us remind that the normalization condition (11) is due to the 7 TeV TOTEM data, which were normalized with a luminosity-dependent method [2]. 
It would be very interesting to compare our predictions with the expected TOTEM data in the whole $ 0.05 < |t| < 3.4\, \mathrm{GeV}^{2}$ region 
at $\sqrt{s} = 13$ TeV. 

\section{Acknowledgements}
We are grateful to V.V.~Ezhela, S.~Giani and J.~Ka\v{s}par for useful discussions.

\end{document}